\documentclass[prd,showkeys,preprint,12pt]{revtex4}
\usepackage{amsmath,amsthm,amsfonts,amssymb}
\usepackage[mathcal]{eucal}
\usepackage{mathrsfs}
\usepackage{graphicx}
\usepackage{dcolumn}
\usepackage{latexsym}
\usepackage{epsfig}
\usepackage{bm}
\usepackage[all]{xy}
\usepackage[utf8]{inputenc}
\usepackage{scalefnt}
\usepackage{subfigure}
\usepackage{subeqnarray}
\usepackage{xcolor}
\usepackage{hyperref}
\hypersetup{
	colorlinks,
    citecolor=blue!20!green!90!black!99,
    filecolor=red,
    linkcolor=blue!20!red!90!black!99,
    urlcolor=gray
    }

\begin{document}

\title{Quasinormal frequencies for a black hole in a bumblebee gravity}
\author{R. Oliveira$^{a}$}
\email{rondinelly@fisica.ufc.br}

\author{D. M. Dantas$^{a}$}
\email{davi@fisica.ufc.br}

\author{C. A. S. Almeida$^{a}$}
\email{carlos@fisica.ufc.br}

\affiliation{$^{a}$Universidade Federal do Cear\'a (UFC), Departamento de F\'{\i}sica, Campus do Pici, Caixa Postal 6030, 60455-760, Fortaleza, Cear\'{a}, Brazil}

\begin{abstract}
After recent observational events like the LIGO-Virgo detections of gravitational waves and the shadow image of M87* supermassive black hole by Event Horizon Telescope (EHT), the theoretical study of black holes was significantly improved.  Quantities as quasinormal frequencies, shadows, and light deflection become more important to analyze black hole models. In this context, an interesting scenario to study is a black hole in the bumblebee gravity. The bumblebee vector field imposes a spontaneous symmetry breaking that allows the field to acquire a vacuum expectation value that generates Lorentz-Violating (LV) into the black hole. In order to compute the quasinormal modes (QNMs) via the WKB method, we obtain the Reege-Wheeler's equation with a bell-shaped potential for this black hole.  Both QNMs, the scalar and tensorial modes, are computed for the black hole in the bumblebee scenario. Moreover, the stability of the Schwarzschild-like solution and some bounds to LV parameters are analyzed.


\end{abstract}
\maketitle

\section{introduction}

The study of black holes is improving its observational knowledge due to the LIGO-Virgo detections
of gravitational waves \cite{Ligo}  and the shadow images of a supermassive black hole M87\text{*} by the international Event Horizon Telescope
(EHT) \cite{tele}. In this context, it is essential to deal with the computation of observable quantities shadows  \cite{Crispino, Kerr-Sen-Bum}, quasinormal modes (QNMs) \cite{KerrQNM} and the deflecting of light \cite{Kerr-Sen def} in black holes and other astronomical compact objects. Moreover, these quantities can help spot modifications in General relativity (GR) and the standard model (SM) of particle physics once that the parameters of modified models should correct the observed result. 

The bumblebee gravity is one of many extensions of the Standard Model Extension (SME) where the Lorentz is spontaneously violated (LV) \cite{KS}. A  Schwarzschild-like solution in the bumblebee gravity is described in the work of Ref. \cite{Casana}. Based on this model, the LV effect over the shadows of a black hole is studied in Ref. \cite{Casana2}, whereas the influence of LV on the black hole accretion is presented in Ref. \cite{Accretion}. Moreover, a wormhole solution in the bumblebee gravity was conceived in the works of Refs. \cite{Ovgun1, Rondinelly}, the radiative corrections in bumblebee electrodynamics are studied in \cite{Euclides} and the black hole with the cosmological constant in bumblebee gravity was presented in Ref. \cite{Maluf}.

The quasinormal frequencies are one of the most frequent quantities for testing black holes. In 1957, Regge and Wheeler computed the QNMs for the Schwarzschild solution \cite{R-W}. After, Zerilli \cite{Zerilli}  extended the work of Regge and Wheeler and described the Schr\"{o}dinger-like equation for these perturbations. The damping oscillations for the QNMs were shown by Vishveshwara \cite{vish}. The approximation method studied by Gregor Wentzel, Hendrik Anthony Kramers, and Leon Brillouin (WKB) of third-order was implemented in the Schwarzschild black hole by S. Iyer and C. Will \cite{Iyer}. Moreover, Konoplya has extended the order of the WKB method for higher-order in reference \cite{Iyer}, where the 6th order approximation has been shown more accurate than the 3rd order.

In this work, we compute the Regge-Wheeler equation for the black hole in the context of bumblebee gravity. The differences of the potentials are detailed for the scalar and gravitational perturbations. For both field perturbations, the  LV parameter slightly modifies the result of the Schwarzschild black hole. Moreover, the quasinormal modes are computed via the WKB method for perturbations to 3rd and 6th orders. Furthermore, the time profile for the scalar and gravitational perturbations are shown. 

\section{QNMs of Schwarzschild-like solution in bumblebee gravity}

A solution of a black hole in the context of the bumblebee gravity was proposed in Ref. \cite{Casana}. The metric of this Schwarzschild-like solution with the Lorentz spontaneously violation is represented by  \cite{Casana}:
\begin{eqnarray}\label{bmetric}
ds^2= \Big(1-\frac{2M}{r}\Big)dt^2 - (1+  \lambda)\left(1-\frac{2M}{r}\right)^{-1} dr^2 - r^2d\theta^2
 - r^2\sin^2\theta d\phi^2,
 \label{Casana}
\end{eqnarray}
where $\lambda  \ll 1$ is the LV parameter (for $ \lambda = 0$ the usual Schwarzschild black hole is obtained).

In order to compute the quasinormal modes (QNM), the WKB method will be applied. Since the solution of Eq. \eqref{Casana} shares some similarities with the Schwarzschild metric, we hope to find a similar QNM  slightly changed by the LV parameter. For this end, the black hole perturbations are described by the Schr\"odinger-like wave equation, the so-called  Regge-Wheeler equation \cite{Iyer, Kono}:
\begin{equation}
    \frac{d^2\, \psi(x)}{dx^2} +  Q(x)\, \psi(x) = 0, \label{RW}
\end{equation}
where $x$ is the tortoise coordinate. The term $Q(x) = \omega^2 - V(x)$ depends on the analogue potential $V(x)$ and on the  frequencies $\omega$. Moreover, the wave function $\psi$ is imposed to following the conditions:
\begin{equation}
    \psi(x) \sim C\pm e^{\mp i\, \omega\, t} \quad \text{com} \quad x \rightarrow \pm \infty.
\end{equation}

Furthermore, the WKB method requires the analogue potential $V(x)$ have some aspects. The potential should have a maximum centered at $x_{0}$, and two returning points $x_{1}$ and $x_{2}$. The potential at limits $(x = -\infty)$ and $(x = +\infty)$ assume constant values. So, the potential should exhibit a bell-shaped profile.

Let us use as an example the usual Schwarzschild black hole. The Schwarzschild solution has the following potential in the original coordinate $r$ \cite{Kono}:
   \begin{equation}
      \mathcal{V}\left(x(r)\right) =  \left(1 - \frac{1}{r}\right) \left( \frac{l(l + 1)}{r^2} + \frac{1-s^2}{r^3}\right), % \quad \text{or} \quad V_s(x)=
      \label{potential-ss}
  \end{equation}
the parameter $l$ is the orbital angular momentum, and $s$ is the spin of field perturbations.

\begin{figure}[!htb]
	\includegraphics[width=.5\linewidth]{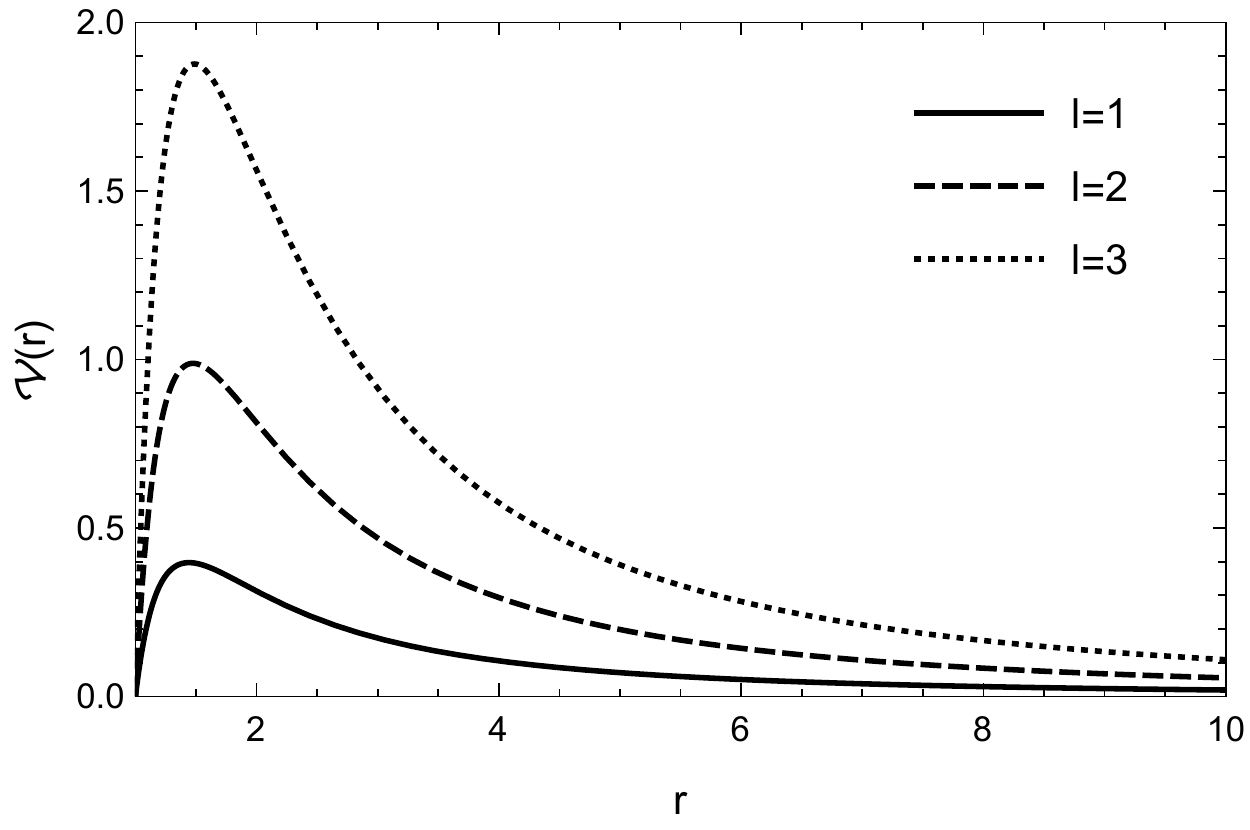}
\caption{The Regge-Wheeler potential for the scalar field in the Schwarzschild black hole with some values of $l$.}
\label{pot-0}
\end{figure}

For $s=0$ the potential for a scalar field is showed in Fig. (\ref{pot-0}), where the bell-shaped behavior is verified. Note that other spins exhibit a similar profile once that the only modification occurs in the term $(1-s^2)$ and the higher perturbation studied here is the gravitational one $(s=2)$.

With these conditions, Iyer and Will \cite{Iyer} found the expression of the QNM via the WKB method as
\begin{equation}
     \frac{i\,Q_{0}}{\sqrt{2\, Q_{0}^{''}}} - \Lambda(n) - \Omega(n) = n + \frac{1}{2},
     \label{WKB3}
\end{equation}
where $Q_{0}$ is the value of the potential at $x_{0}$ and the primes denote derivatives concerning $x$ coordinates. Corrections compose the terms $\Lambda(n)$ and $\Omega(n)$ up to third order \cite{Rondinelly, Iyer}. An improved version of WKB method was presented in Ref. \cite{Kono}, where high order corrections were studied. For the sixth order, the WKB method can be written as \cite{Kono}
 \begin{equation}
     \frac{i\,Q_{0}}{\sqrt{2\, Q_{0}^{''}}} - \Lambda_{2}(n) - \Lambda_{3}(n) - \Lambda_{4}(n) -\Lambda_{5}(n) - \Lambda_{6}(n) = n + \frac{1}{2}.
     \label{WKB6}
 \end{equation}

 The Tables \ref{table-kono-1} e \ref{table-kono-2} are the results of Ref. \cite{Kono} of the WKB method (third and sixth orders) for the scalar and gravitational perturbations in Schwarzschild black hole. Note that the results for the 6th order are better approximated to the numerical values than the 3rd order for the WKB method.

\begin{table}[!htb]
	\begin{tabular}{|c|c|c|c|c|}
		\hline
		$ l, \ n $  & numerical    & 3rd order WKB     & 6th order WKB  \\  
		\hline\hline
%$l = 0, n = 0$& $0.1105 - 0.1049i$& $0.1046 - 0.1152i$&  $0.1105 - 0.1008i$ \\ \hline
$l = 1, n = 0$& $0.2929 - 0.0977i$& $0.2911 - 0.0980i$&  $0.2929 - 0.0977i$ \\ \hline
%$l = 1, n = 1$& $0.2645 - 0.3063i$& $0.2622 - 0.3074i$& $0.2645 - 0.3065i$ \\  \hline
$l = 2, n = 1$& $0.4639 - 0.2956i$& $0.4632 - 0.2958i$& $0.4638 - 0.2956i$\\  \hline
$l = 2, n = 2$& $0.4305 - 0.5086i$& $0.4317 - 0.5034i$& $0.4304 - 0.5087i$  \\  \hline
	\end{tabular}
	\caption{The QNM for $s = 0$, results from Ref. \cite{Kono}.}\label{T1}
	\label{table-kono-1}
\end{table}

\begin{table}[!htb]
	\begin{tabular}{|c|c|c|c|c|}
		\hline
		$ l, \ n $  & numerical    & 3rd order WKB     & 6th order WKB  \\  
		\hline\hline
$l = 2, n = 0$& $0.3737 - 0.0890i$& $0.3732 - 0.0892i$&  $0.3736 - 0.0890i$ \\ \hline
$l = 2, n = 1$& $0.3467 - 0.2739i$& $0.3460 - 0.2749i$&  $0.3463 - 0.2735$ \\ \hline
$l = 2, n = 2$& $0.3011 - 0.4783i$& $0.3029 - 0.4711i$& $0.2985 - 0.4776i$ \\  \hline
$l = 3, n = 0$& $0.5994 - 0.0927i$& $0.5993 - 0.0927i$& $0.5994 - 0.0927i$\\  \hline
$l = 3, n = 1$& $0.5826 - 0.2813i$& $0.5824 - 0.2814i$& $0.5826 - 0.2813i$  \\  \hline
$l = 3, n = 2$& $0.5517 - 0.4791i$& $0.5532 - 0.4767i$& $0.5516 - 0.4790i$ \\  \hline
$l = 3, n = 3$& $0.5120 - 0.6903i$& $0.5157 - 0.6774i$& $0.5111 - 0.6905i$ \\  \hline
	\end{tabular}
	\caption{The QNM for $s = 2$, results from Ref. \cite{Kono}.}\label{T2}
	\label{table-kono-2}
\end{table}

 In this work, we compute the Regge-Wheeler equation of Eq. \eqref{RW} using the metric given in Eq. \eqref{Casana}, for both perturbations of the scalar (spin $s = 0$) and the gravitational (spin $s = 2$) fields. As a result, the quasinormal modes  will be obtained from eq. \eqref{WKB6}, and the influence of the bumblebee parameter over frequencies will be analyzed. Also, we compare the differences in the 3rd and 6th order of the WKB method.

 \section{Quasinormal modes of perturbations of scalar fields}
 
 In this section, the first case of spin $s=0$ scalar field will be described. The massless Klein-Gordon equation in the presence of gravity leads to the following equation of motion \cite{Gordon}
 \begin{equation}
     \frac{1}{\sqrt{- g}}\partial_{\mu}(g^{\mu\nu}\sqrt{- g} \,\partial_{\nu}\, \Phi) = 0.
 \end{equation}
 Applying the coordinates decomposition, the variables can be separated as follows:
 \begin{equation}
 \Phi = \sum_{l = 0}^{\infty}\sum_{m = -l}^{l}\frac{R(r, t)}{r}\,Y_{l\, m}(\theta, \phi),
 \label{Variaveis}
  \end{equation}
where $Y_{l\, m}(\theta, \phi)$  are the spherical harmonics.

By substituting this decomposition of Eq. \eqref{Variaveis} into Eq. \eqref{RW}  with the metric of Eq. \eqref{Casana} we obtain:
  \begin{equation}
        \frac{d^2\, \phi(x)}{dx^2} +  (\omega^2 - V_{s}(l, x, \lambda))\, \phi(x) = 0,
        \label{pot-escalar}
  \end{equation}
where for Eq. \eqref{pot-escalar} the transformation of tortoise coordination $x$ is given by
  \begin{equation}
      \frac{dx}{dr} = \frac{\sqrt{1 + \lambda}}{ \left(1 - \frac{1}{r}\right)}, \quad \Rightarrow \quad x(r)=\sqrt{1 + \lambda}\left[r+ \ln(r-1)\right].
      \label{torte}
  \end{equation}

The potential of Regge-Wheeler for the scalar field in the context of bumblebee gravity has the formula
  \begin{equation}
      V_{s}(r) =  \left(1 - \frac{1}{r}\right)\left( \frac{l(l + 1)}{r^2} + \frac{\left[1 + \lambda\right]^{-1}}{r^3}\right).
      \label{potential-s}
  \end{equation}
  
\begin{figure}[!htb]
	\includegraphics[width=.5\linewidth]{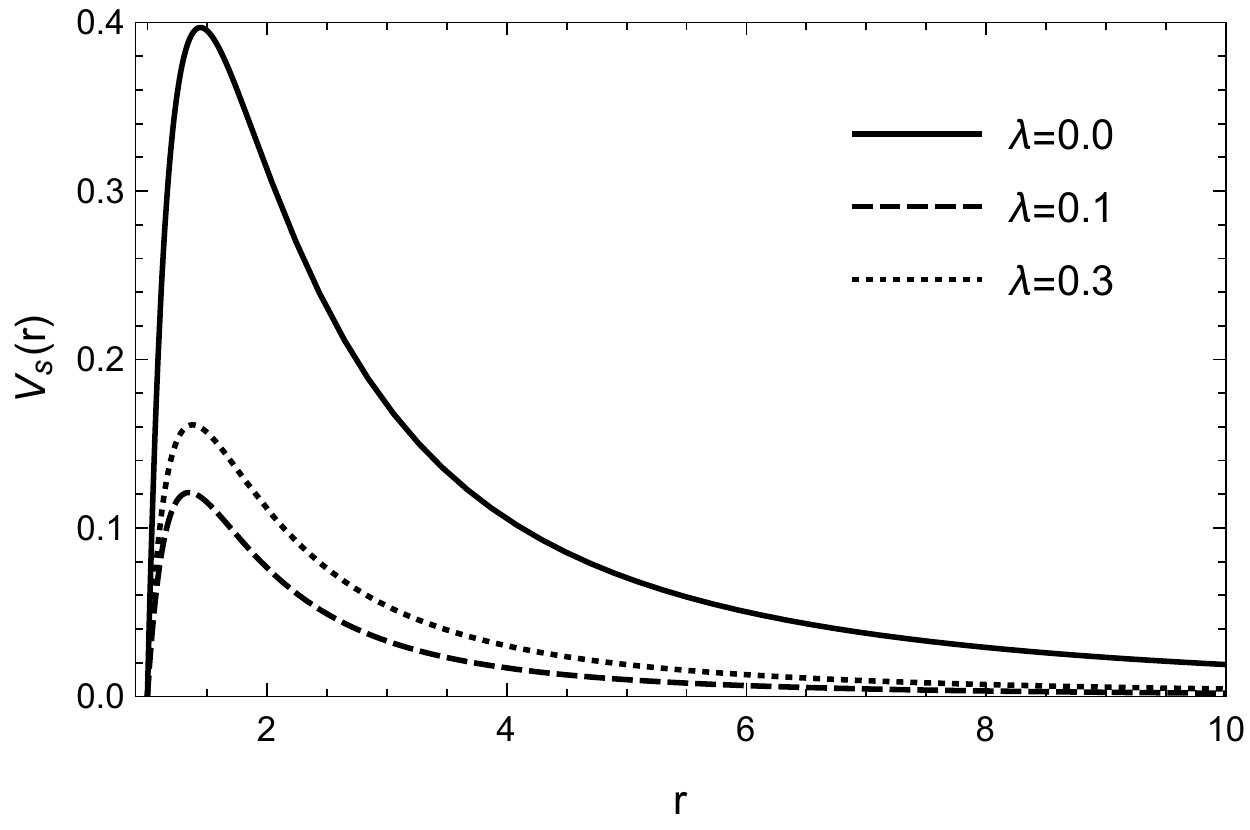}
\caption{Potential for scalar field in the bumblebee black hole with some values of $\lambda$.}
\label{pot-lambda}
\end{figure}

Note that by Eq. \eqref{potential-s}, the bumblebee LV parameter $(\lambda)$ modifies the potential of the usual Schwarzschild black hole in Eq. \eqref{potential-ss} by the term $\left[1 + \lambda\right]^{-1}$. As expected, for $\lambda=0$ the usual potential is obtained. The LV parameter decreases the peaks of the potential, as can be shown in Fig. (\ref{pot-lambda}) for $l=1$.

By inverting $r(x)$ in Eq. \eqref{torte} and obtained the $V(x)$ in Eq. \eqref{potential-s}, the terms of WKB of 3rd order in Eq. \eqref{WKB3} and 6th order in Eq. \eqref{WKB6} can be computed. The results of QNM for scalar perturbation in bumblebee gravity are shown in Table \ref{T3}, for the 3rd and 6th order, respectively. It is worth highlighting that $n \leq l$ needs to be imposed.
 
\begin{table}
	\begin{tabular}{|c|c|c|c|c|}
		\hline
		$ l\, ,\, n$  &  3rd order WKB   & 6rd order WKB \\  
		\hline\hline
		l = 1, n = 0& 0.291644 - 0.089100i&  0.289044 - 0.097382i \\  \hline
		l = 2, n = 0& 0.483357 - 0.088035i& 0.481314 - 0.096615i \\  \hline
		l = 2, n = 1& 0.466639 - 0.268196i & 0.461409 - 0.295206i\\  \hline
		l = 3, n = 0& 0.675267 - 0.087756i& 0.673701 - 0.096422i \\  \hline
		l = 3, n = 1& 0.662979 - 0.265386i& 0.658965 - 0.292061i \\  \hline
		l = 3, n = 2& 0.641291 - 0.447758i & 0.63181 - 0.495662i\\ \hline
		l = 4, n = 0& 0.867373 - 0.087646i & 0.866120 - 0.096344i\\ \hline
		l = 4, n = 1& 0.857714 - 0.264225i & 0.854492 - 0.290737i  \\ \hline
		l = 4, n = 2& 0.839892 - 0.444019i & 0.801780 - 0.697192i\\ \hline
		l = 4, n = 3& 0.815917 - 0.627569i & 0.765748 - 0.913890i\\ \hline
		l = 5, n = 0& 1.05960 - 0.0877264i & 1.05961 - 0.096337i \\ \hline
		l = 5, n = 1& 1.05183 - 0.263979i & 1.050040 - 0.290154i\\ \hline
		l = 5, n = 2& 1.03709 - 0.442350i & 1.03150 - 0.487343i \\ \hline
		l = 5, n = 3& 1.01659 - 0.623493i & 1.005170 - 0.689918i\\ \hline 
	\end{tabular}
	\caption{The QNMs for scalar perturbations in bumblebee gravity via the WKB method for $\lambda = 0.1$}\label{T3}
\end{table}

\begin{figure}[!htb]
\includegraphics[width=.5\linewidth]{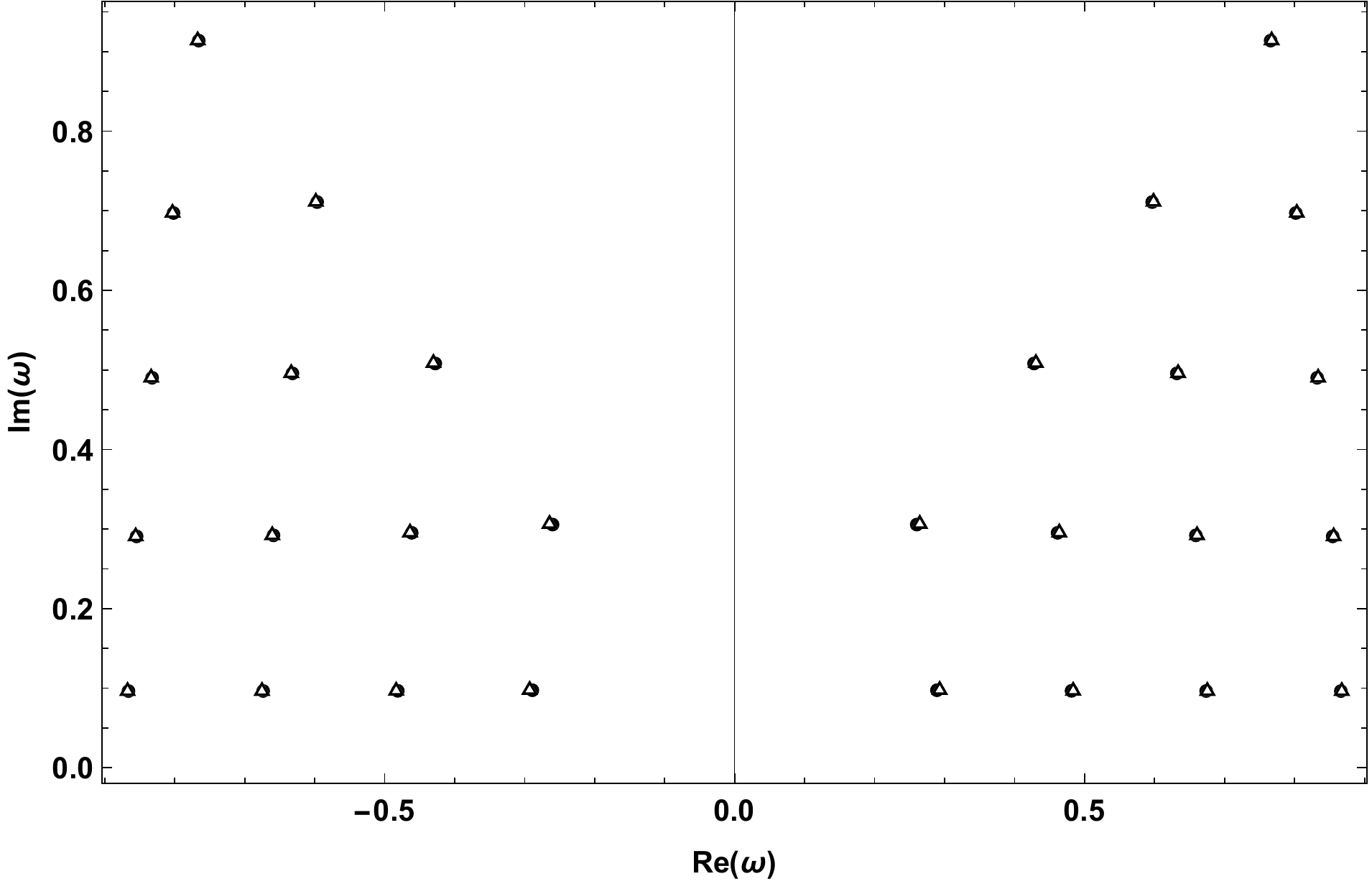}
\caption{The QNM of 6th order for spin $s = 0$  with  $l = 2, 3, 4, 5$. The model with bumblebee gravity is represented by circles where $\lambda=0.1$. The Schwarzschild black hole $(\lambda =0)$ is represented by triangles.}
\label{com}
\end{figure}

The QNM for the scalar field in bumblebee gravity is also denoted in Fig. (\ref{com}). Note the differences between our results, presented in Table \ref{T1}  (denoted by circles) with the results for the usual Schwarzschild solution in the same Table \ref{table-kono-1}, but denoted by triangles.  From Fig. (\ref{com}) we conclude that the LV parameter slightly decreases for the real part as well as for the imaginary part of frequencies.

In the next section, we apply the same methodology for the gravitational perturbations ($s=2$), where we conclude the results are similar to the scalar perturbation.

 \section{Quasinormal modes of gravitational field perturbations}
 
 In order to study the gravitational perturbations, the Chandrasekhar formalism \cite{chan} was applied. As expected, in the same way as scalar perturbations, the Regge-Wheeler equation is similar to the Schwarzschild case one. The Regge-Wheeler potential in the context of bumblebee gravity reads
 \begin{equation}
     V_g(r, l, \lambda) = \left(1 - \frac{1}{r}\right) \left(\frac{l(l + 1)}{r^2} -\frac{2}{r^2} +\left[1 + \lambda\right]^{-1}\left[\frac{2}{r^2}-\frac{3}{r^3}\right]\right).
     \label{potential-g}
 \end{equation}

Note  the gravitational potential in Eq. \eqref{potential-g} differs from the gravitational potential $(s=2)$ for Schwarzschild solution in Eq.  \eqref{potential-ss} by the term $\left[1+\lambda\right]^{-1}\left[\frac{2}{r^2}-\frac{3}{r^3}\right]$, which for $\lambda=0$ recovers the usual result.

\begin{figure}[htb]
	\includegraphics[width=.5\linewidth]{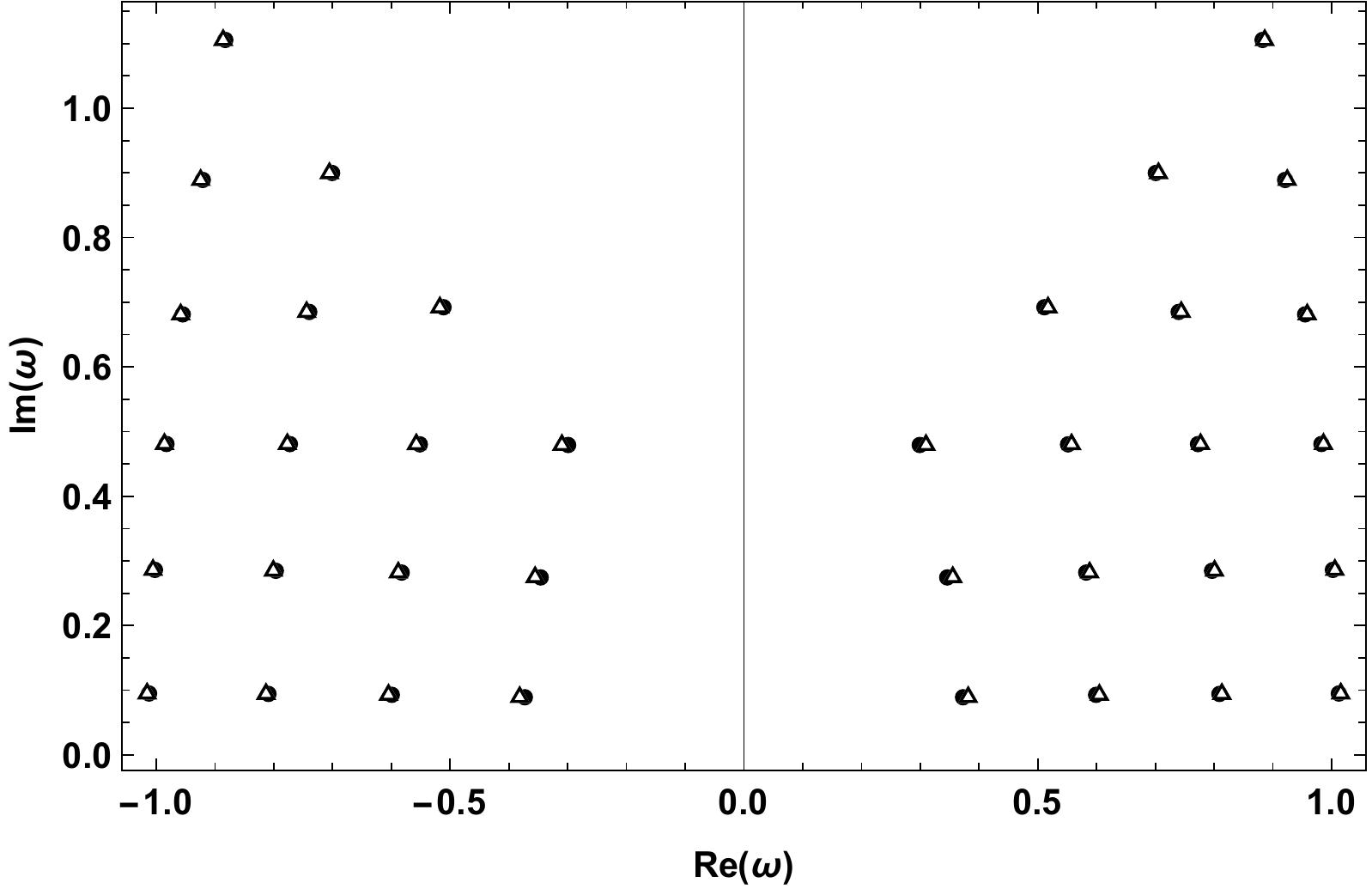}
\caption{The QNMs of 6th order for spin $s = 2$  with  $l = 2, 3, 4, 5$. The model with bumblebee gravity is represented by circles where $\lambda=0.1$. The Schwarzschild black hole $(\lambda =0)$ is represented by triangles.}
\label{graf}
\end{figure}

The same transformation $x(r)$ for the tortoise coordinate presented in Eq. \eqref{torte} and same constructions for the the QNMs of \eqref{WKB3} and \eqref{WKB6}, leads us to the new QNM for the gravitational field which is presented in Table \ref{T3} for $\lambda=0.1$. Once more, we compare our results in the bumblebee gravity (Table \ref{T4}) with the Schwarzschild black hole (\ref{table-kono-2}) in Fig. (\ref{graf}), now for the gravitational perturbations. As also verified for scalar perturbations, the LV slightly modifies the QNMs of Schwarzchild scenario.

\begin{table}
	\begin{tabular}{|c|c|c|c|}
		\hline
		$ l\, ,\, n$  &  3rd orde WKB   & 6rd orde WKB      \\  
		\hline\hline
		l = 2, n = 0& 0.375195 - 0.081563i& 0.372867 - 0.0891936i \\  \hline
		l = 2, n = 1& 0.352768 - 0.250222i& 0.345894 - 0.274473i \\  \hline
		l = 3, n = 0& 0.60087 - 0.0846012i & 0.599221 - 0.092926i\\  \hline
		l = 3, n = 1& 0.586885 - 0.25624 i& 0.582443 - 0.281986i \\  \hline
		l = 3, n = 2& 0.562304 - 0.433172i& 0.551488 - 0.480268i \\  \hline
		l = 4, n = 0& 0.810376 - 0.0857977i& 0.809084 - 0.094306i\\ \hline
		l = 4, n = 1& 0.799958 - 0.258791i& 0.796545 - 0.284766i\\ \hline
		l = 4, n = 2& 0.780779 - 0.435239i & 0.772636 - 0.480641i  \\ \hline
		l = 4, n = 3& 0.755005 - 0.615642i& 0.739674 - 0.684973i\\ \hline
		l = 5, n = 0& 1.01331 - 0.0863744i & 1.01225 - 0.094966i\\ \hline
		l = 5, n = 1& 1.00495 - 0.260035i & 1.00218 - 0.286108i \\ \hline
		l = 5, n = 2& 0.989174 - 0.436100i& 0.982659 - 0.480820i\\ \hline
		l = 5, n = 3& 0.967365 - 0.615274i & 0.954953 - 0.681248i \\ \hline
		l = 6, n = 0& 1.21288 - 0.0866971i & 1.01225 - 0.0949664i\\ \hline
		l = 6, n = 1& 1.20589 - 0.260735i & 1.00218 - 0.286108i \\ \hline
	\end{tabular}
	\caption{The QNMs for gravitational perturbations in bumblebee gravity via the WKB method for $\lambda = 0.1$
}\label{T4}
\end{table}

\begin{figure}[!htb]
\begin{minipage}{0.47\textwidth}
    \centering
    \includegraphics[width=.99\linewidth]{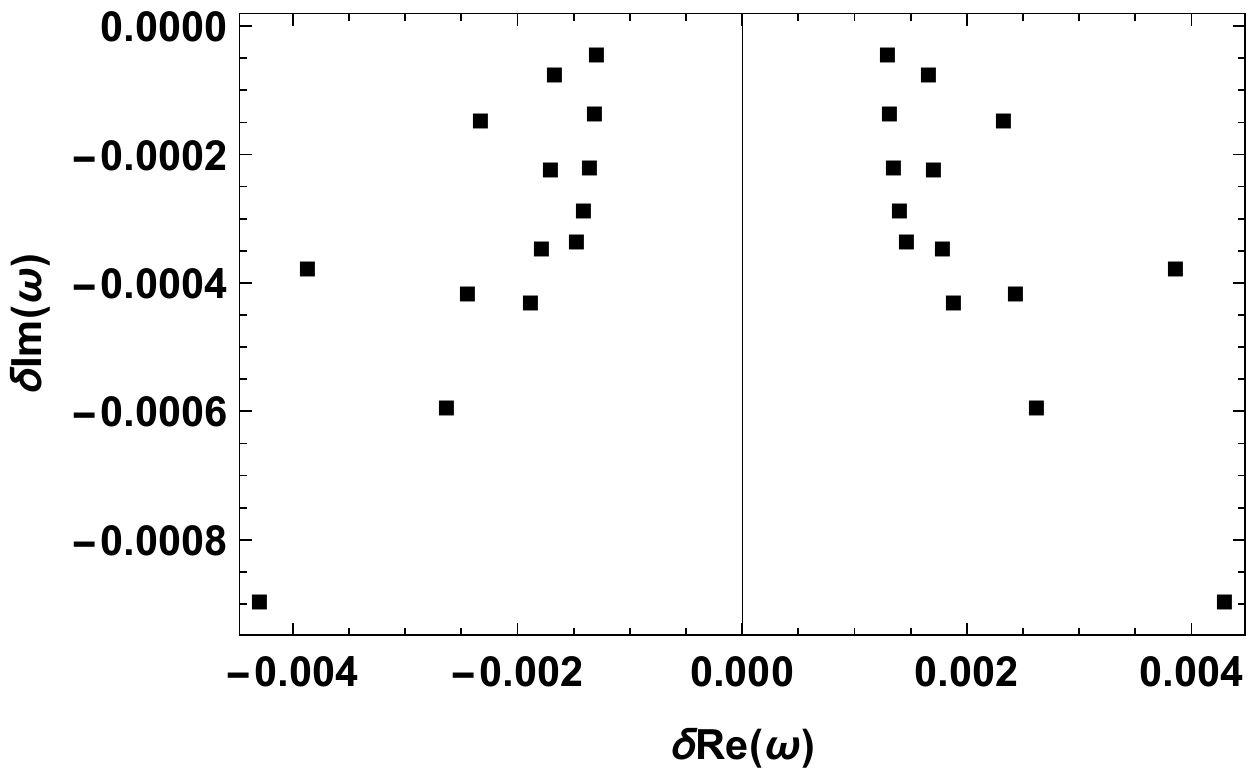}
     \centering
%\caption{Scalar dumping}
%\label{dumping10}
    \end{minipage}%
    \qquad
    \begin{minipage}{0.47\textwidth}
        \centering
        \includegraphics[width=.99\linewidth]{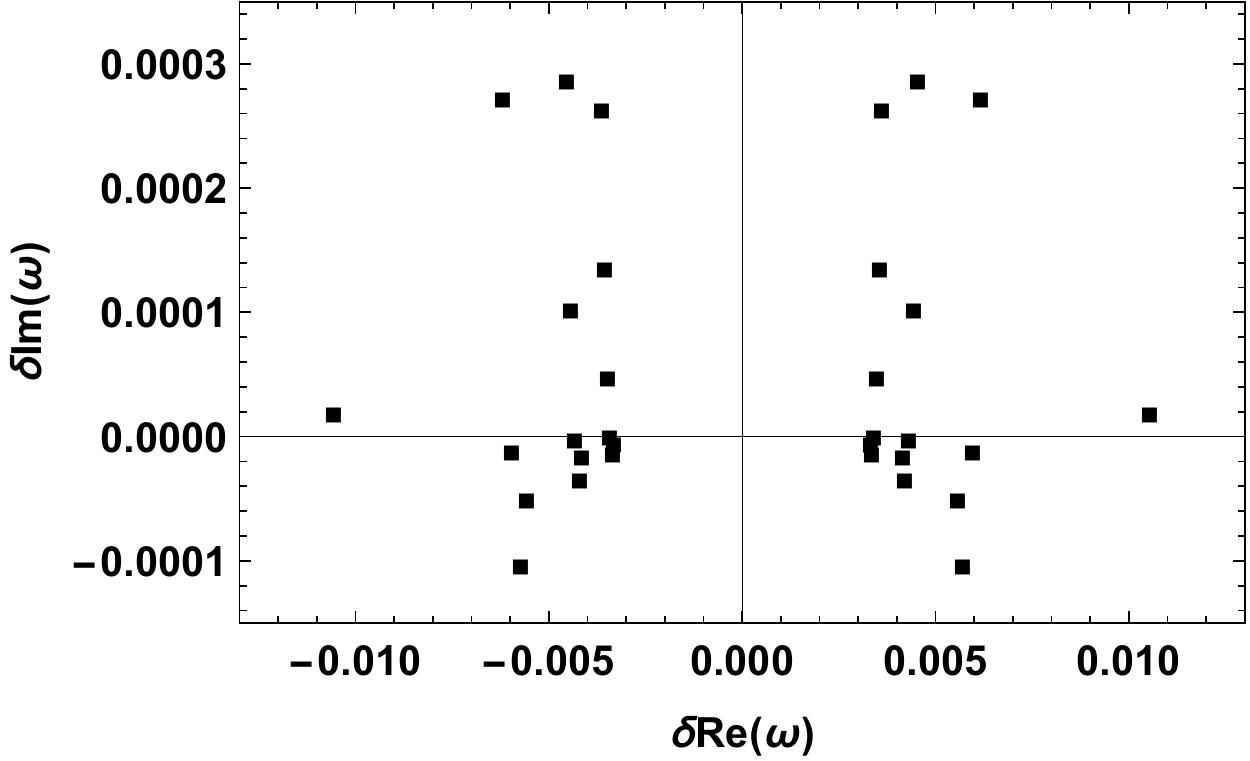}
        \end{minipage}
        \caption{Variations of QNMs frequencies via the WKB method for the 6th order. The scalar perturbation is shown in the left panel, and the gravitational perturbation in the right panel.}
\label{diff}
\end{figure}

In order to better spot the QNM differences between the bumblebee gravity and the Schwarzschild black hole, we plot in Fig. (\ref{diff}) only the variations of QNM for the 6th order. Note that, for scalar perturbations, the corrections always represent an increasing in the real part and a decrease in the imaginary part of frequencies. For gravitational perturbations, the corrections for the imaginary part can be positive or negative.

%\section{Time domain}

\begin{figure}[!htb]
\begin{minipage}{0.47\textwidth}
    \centering
    \includegraphics[width=.99\linewidth]{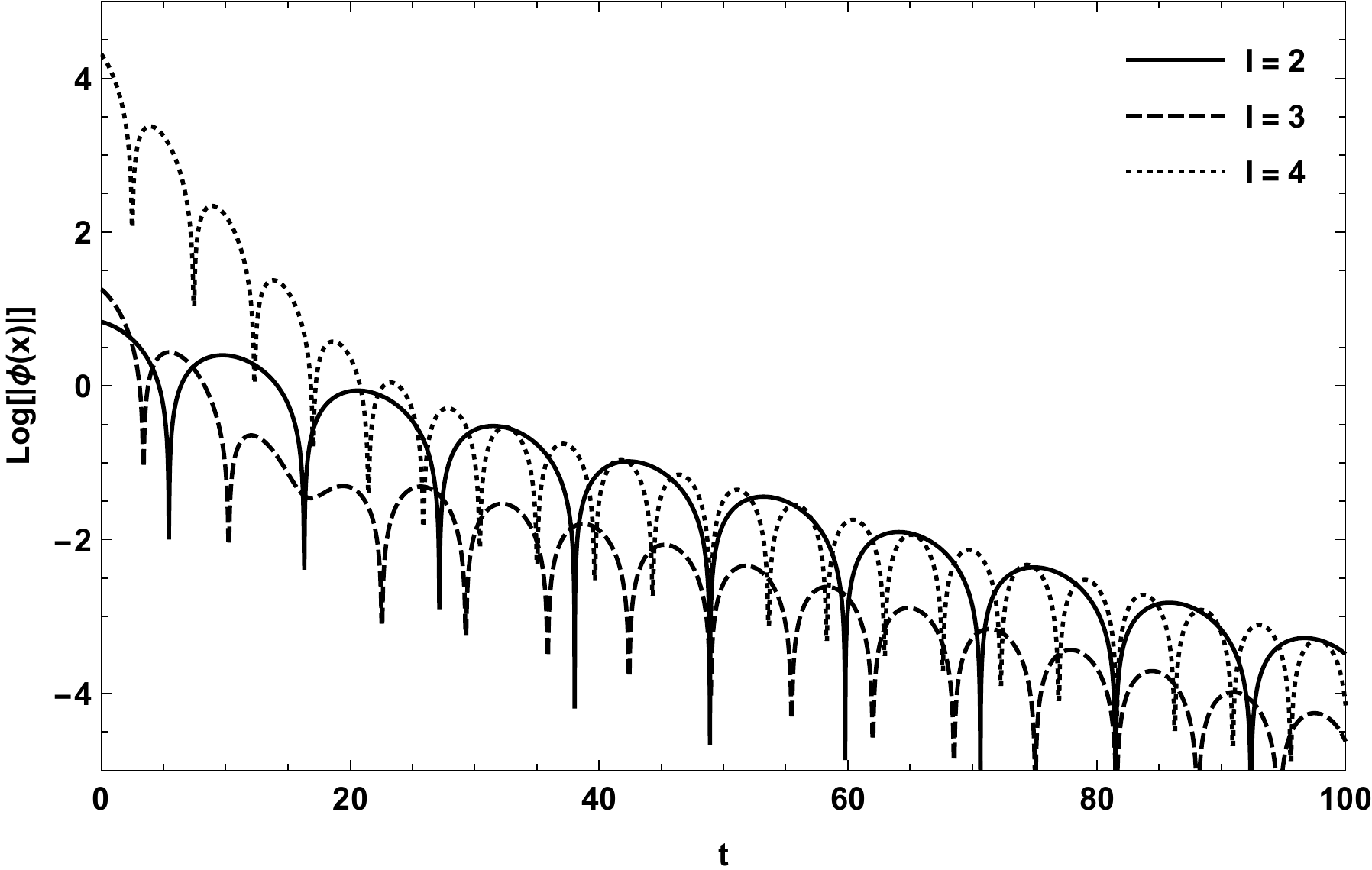}
     \centering
%\caption{Scalar dumping}
%\label{dumping10}
    \end{minipage}%
    \qquad
    \begin{minipage}{0.47\textwidth}
        \centering
        \includegraphics[width=.99\linewidth]{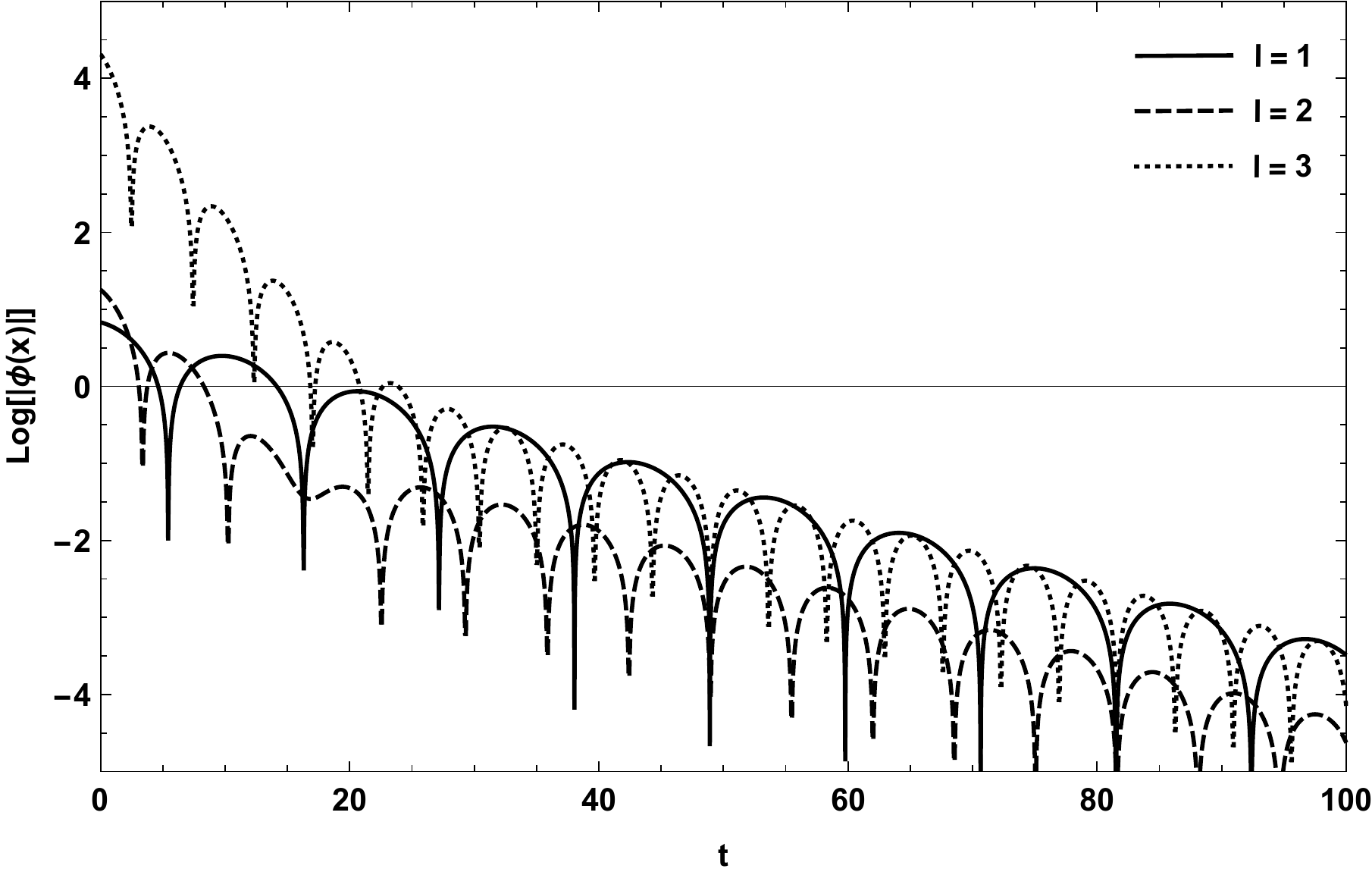}
        \end{minipage}
        \caption{Scalar damping (left panel) and Gravitational damping (right panel) for $\lambda=0.1$ and some values of $l$.}
\label{dumping10}
\end{figure}

Moreover, in Fig. (\ref{dumping10})  the time domain for scalar and gravitational perturbation in the black hole with bumblebee parameter is exhibited. Note that both perturbations exhibit damping profiles, decreasing scalar modes slower than the gravitational modes. The angular momentum $l$ parameter faster the decay of oscillations as verified in references \cite{Rondinelly, kono1, chur}.

\section{Conclusion}

In this work, we apply the WKB method to compute the quasinormal modes for a black hole in bumblebee gravity, and the  results were compared to the Schwarzschild black hole.  The scalar and gravitational perturbations were considered. The Regge-Wheeler equation was obtained, and the correction term responsible for the Lorentz violation was addressed. Once the modified potential still exhibits a bell-shaped, the WKB method was applied for the 3rd and 6th order. The LV parameter generated small variations in relation to QNMs of Schwarzschild solution. Moreover, the time domain for both perturbations was analyzed and represent damping profiles.

\section*{Acknowledgments}
\hspace{0.5cm}The authors thank the Conselho Nacional de Desenvolvimento Cient\'{\i}fico e Tecnol\'{o}gico (CNPq), grant n$\textsuperscript{\underline{\scriptsize o}}$ 308638/2015-8 (CASA), and Coordena\c{c}\~{a}o de Aperfei\c{c}oamento do Pessoal de N\'{i}vel Superior (CAPES), for financial support.

\section{bibliography}

\end{document}